\documentclass[conference]{IEEEtran}
\IEEEoverridecommandlockouts
\usepackage{cite}
\usepackage{amsmath,amssymb,amsfonts}
\usepackage{graphicx}
\usepackage{xcolor}
\usepackage{textcomp}
\usepackage{booktabs}
\usepackage{multirow}
\usepackage{url}
\usepackage{amsthm}
\theoremstyle{definition}
\newtheorem{lemma}{Lemma}
\graphicspath{{figs/}}
\newcommand{\figref}[1]{Fig.~\ref{#1}}
\newcommand{\tabref}[1]{Table~\ref{#1}}

\begin{document}

\title{DGEMM with Ozaki Scheme I/II on FP4 Tensor Cores: A Base-13 E2M1 Limb Representation}

\makeatletter
\newcommand{\linebreakand}{%
  \end{@IEEEauthorhalign}
  \hfill\mbox{}\par
  \mbox{}\hfill\begin{@IEEEauthorhalign}
}
\makeatother

\author{
\IEEEauthorblockN{Shun-ichiro Hayashi}
\IEEEauthorblockA{Graduate School of Informatics\\
Nagoya University\\
Aichi, Japan\\
hayashi@hpc.itc.nagoya-u.ac.jp}
\and
\IEEEauthorblockN{Daichi Mukunoki}
\IEEEauthorblockA{Information Technology Center\\
Nagoya University\\
Aichi, Japan\\
mukunoki@cc.nagoya-u.ac.jp}
\linebreakand
\IEEEauthorblockN{Tetsuya Hoshino}
\IEEEauthorblockA{Information Technology Center\\
Nagoya University\\
Aichi, Japan\\
hoshino@cc.nagoya-u.ac.jp}
\and
\IEEEauthorblockN{Takahiro Katagiri}
\IEEEauthorblockA{Information Technology Center\\
Nagoya University\\
Aichi, Japan\\
katagiri@cc.nagoya-u.ac.jp}
}

\maketitle

\begin{abstract}
This paper proposes a method and its implementation for emulating FP64
matrix multiplication (DGEMM) by constructing, on FP4 (E2M1; 2 exponent
bits and 1 mantissa bit) Tensor Cores, Ozaki schemes I and II, which
realize high-precision matrix multiplication on low-precision arithmetic
units. Prior implementations were based on INT8 and FP8, and the use of
the faster FP4 had not been realized. The key property is that
every FP4 value becomes an integer when doubled, and that shifting
this integer set by multiples of 13 covers all integers. Converting an arbitrary integer into
base-13 FP4 limbs by this property keeps intermediate sums error-free in FP32 accumulators, which makes FP4 Tensor Cores usable for Ozaki
schemes I and II. By the same principle, the integer GEMM of INT8 Tensor
Cores can also be emulated bit-exactly on FP4 Tensor Cores. When FP4 Tensor
Cores have twice the throughput of FP8, Ozaki scheme II on FP4
theoretically achieves slightly higher performance than its FP8
counterpart. This paper further proposes kernel implementation
optimizations raising the attained fraction of peak performance, obtaining a measured speedup on top of the theoretical advantage. We verify
this on an RTX PRO 6000 Blackwell, achieving performance competitive
with that of an existing FP8-based implementation of Ozaki scheme II, and
actually exceeding it at a large problem size ($16384^3$).
\end{abstract}


\section{Introduction}
\label{sec:intro}

Driven by the rapid spread of large language models (LLMs), GPU vendors
preferentially allocate their limited silicon area to matrix
multiply-accumulate (MMA) units of very low precision. Meanwhile, the FP64
units that have supported scientific computing are being scaled back. On
the B300 (Blackwell Ultra, 2025), NVIDIA's latest GPU at the time of
writing, the FP64 units were reduced by more than 90\,\% from the previous generation~\cite{nvidia2025blackwellultra}. This is not specific
to the B300: on the next-generation Rubin as well, native FP64 will remain
at the level of the Hopper generation
(2022)~\cite{register2026fp64emu}. There is therefore concern that
performance improvements in scientific computing that requires FP64 will stagnate. This has motivated research on emulating high-precision
arithmetic such as FP64 with high-throughput low-precision MMA units.

This paper targets the emulation of FP64 matrix multiplication (DGEMM),
a representative kernel of scientific computing: $C = AB$ with
$A, B \in \mathbb{F}^{N \times N}$, where
$\mathbb{F} \subset \mathbb{R}$ is the set of numbers representable in
FP64. Each
element $(AB)_{ij} = \sum_{k} A_{ik} B_{kj}$ is in general a real number
not in $\mathbb{F}$; with $\mathrm{fl} : \mathbb{R} \to \mathbb{F}$
denoting round-to-nearest, the ideal output is $C = \mathrm{fl}(AB)$, where each element is rounded exactly once. Matrix multiplication is
compute-bound, requiring only $O(N^2)$ data movement for $O(N^3)$
arithmetic, so its performance is determined by the throughput of the
arithmetic units.

The basis for this emulation should be whichever low-precision format has
the highest throughput at the time. Arithmetic precision has stepped down
from FP16/BF16 to FP8 and then to FP4 (E2M1; 2 exponent bits and 1
mantissa bit), raising throughput at each step. FP8 was introduced in the
Hopper generation, and the fifth-generation Tensor Cores of the following
Blackwell generation (2024) introduced FP4 with twice the throughput
of FP8.
FP4 was adopted in 2023 as MXFP4 in the Open Compute Project (OCP)
Microscaling (MX) standard~\cite{ocp2023mx}, an industry standard sharing one scale per block. NVIDIA's NVFP4~\cite{nvidia2025nvfp4} is a
variant that keeps the element format E2M1 while raising quantization
accuracy with smaller blocks and higher-precision scales. On the
application side as well, the bytes per weight are halved relative to FP8,
doubling effective throughput, and NVFP4 keeps the accuracy loss under
1\,\%; major models such as DeepSeek are accordingly distributed in
FP4-quantized form, and the use of FP4 is spreading beyond inference
quantization into training.

The representative framework for realizing high-precision matrix
multiplication on low-precision matrix units is the Ozaki scheme. It has
two generations: the original, based on splitting the mantissa into digits
(Ozaki scheme I~\cite{ozaki2012eftmm}), and a second generation (Ozaki
scheme II~\cite{ozaki2025schemeii}) based on a residue number system
(RNS) and the Chinese remainder theorem (CRT). Its implementations have
progressed from FP16 to INT8 and FP8 units. Since INT8 units were sharply
reduced on B300 and next-generation GPUs are expected to follow the same design favoring low-precision floating point, FP8 is expected to serve
as the basis for the Ozaki scheme. FP4, however, with twice the throughput of even
FP8, was passed over in prior work~\cite{uchino2026fp8ozakiii} on the
grounds that intermediate sums cannot be held error-free within the
format, and no implementation has used it as a basis.

This paper proposes the use of FP4 Tensor Cores in the Ozaki scheme. The
principle is to exploit the integer nature of FP4 and rewrite an
arbitrary integer in a positional base-13 representation. Because the
intermediate sums are kept in the integer range of the FP32 accumulators
rather than written back to the FP4 format, the proposed method avoids the
problem that led prior work to pass over FP4. Building on this
representation, both Ozaki schemes I and II can be adapted to FP4, and on
hardware where FP4 has twice the throughput of FP8, the FP4 version of
Ozaki scheme II theoretically outperforms the FP8 version slightly even after accounting for the increased number of GEMMs. In
addition to this theoretical advantage, this paper proposes kernel
implementation
optimizations that raise the attained fraction of peak performance, and
demonstrates by measurement on an RTX PRO 6000 Blackwell that the method
is competitive with the existing FP8-based implementation and, at a large
problem size ($16384^3$), outperforms it even end to end, including
preprocessing.

The contributions of this paper are as follows.
\begin{enumerate}
\item A formalization of the integer nature of FP4 and of the fact that
shifting its doubled value set by multiples of 13 covers all integers, and the
demonstration that FP4 Tensor Cores alone can thereby emulate the integer
GEMM of INT8 Tensor Cores (INT8$\times$INT8$\to$INT32) bit-exactly
(Section~\ref{sec:base}).
\item Constructions of both Ozaki schemes I and II on FP4 based on this
principle. Ozaki scheme II in particular requires a newly designed set of
moduli whose residues are representable in FP4, which we provide
(Sections~\ref{sec:ozaki1} and~\ref{sec:ozaki2}).
\item Performance competitive with that of GEMMul8~\cite{riken2026gemmul8},
the existing implementation of FP8 Ozaki scheme II, and superior to it at
problem size $16384^3$. This advantage is the product of a theoretical difference
stemming from the FP4:FP8 throughput ratio and the GEMM-count ratio, and a
higher attained fraction of peak performance due to kernel implementation
optimizations; we explain it with a runtime model
(Sections~\ref{sec:duel} and~\ref{sec:model}).
\end{enumerate}

\section{Related Work}
\label{sec:related}

Ozaki scheme I~\cite{ozaki2012eftmm} was proposed as an error-free transformation decomposing a matrix product into several products
without error, originally as a method for computing matrix products to
high accuracy. Implementations using it to emulate high-precision
matrix multiplication with low-precision GEMMs followed, moving from FP16
Tensor Cores~\cite{mukunoki2020tensorcore} to INT8 units
(ozIMMU)~\cite{ootomo2024dgemm,ootomo2024ozimmu}
with a subsequent reduction in the number of
products~\cite{uchino2025enhancement}, and to FP8 Tensor
Cores~\cite{mukunoki2026fp8}. In contrast to the slices of these prior
studies (components of the mantissa cut out at fixed widths along bit
boundaries),
the proposed method uses an arithmetic representation that rewrites
integers in base 13, without splitting at bit boundaries.

Ozaki scheme II~\cite{ozaki2025schemeii} is a separate approach that uses
a residue number system with CRT reconstruction; it was followed by an
INT8 Tensor Core implementation~\cite{uchino2025power}, an FP8-quantized
version~\cite{uchino2026fp8ozakiii}, and the public implementation
GEMMul8~\cite{riken2026gemmul8}.
The FP8 version of GEMMul8 splits the residue for each modulus into two
parts and reduces the number of partial products per modulus from four to
three with Karatsuba's method and related techniques. The proposed method
can be positioned as a variant replacing the ``low-precision digits''
of both schemes with the base-13 FP4 representation. The FP8 Ozaki scheme
II paper~\cite{uchino2026fp8ozakiii} did consider FP4 as a candidate
basis, but rejected it because the partial sums that arise in computing
the residues for each modulus are inevitably rounded when written back to
the FP4 format, with its coarse 1-bit mantissa and narrow range;
consequently, intermediate sums cannot be held error-free within the FP4
format. The
same paper also projected that an FP4 configuration would become
advantageous only when FP4 throughput exceeds three times that of FP8.
This work avoids the former by keeping the intermediate sums in the
integer range of the FP32 accumulators instead of returning them to the
FP4 format. As for the latter, our choice of moduli holds the number of
GEMMs to 75 and lowers the required throughput ratio to about 1.9.

On the kernel-design side, EmuGEMM~\cite{lu2026emugemm}, a fused
Tensor Core kernel that eliminates round trips of intermediate data
through global memory, addresses the same concern as our kernel fusion.
NVIDIA has also integrated ADP~\cite{schwarz2026adp}, which automatically
selects the number of slices to guarantee DGEMM accuracy, into cuBLAS,
positioning this kind of emulation as a means of compensating for
insufficient FP64 performance~\cite{register2026fp64emu}.

\section{Base-13 FP4 Limb Representation}
\label{sec:base}

\begin{figure*}[tb]
\centering
\includegraphics[width=0.65\linewidth]{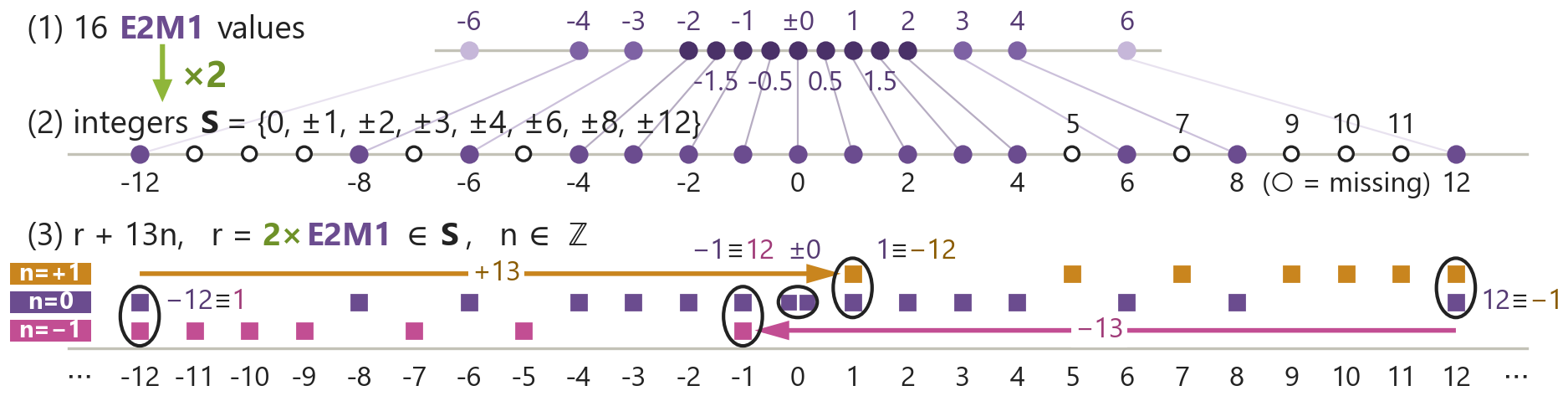}
\caption{The E2M1 value distribution (top) is doubled into integers
(middle); $r + 13n$ ($r$ is a doubled E2M1 value with $r \in S$, $n$ an
integer; the figure shows $|n| \le 1$) represents all integers in
$[-12, 12]$ (bottom). Rings mark redundant representations ($\pm 0$,
$1 = -12 + 13$, etc.).}
\label{fig:coverage}
\end{figure*}

For both Ozaki schemes I and II, the proposed method converts the inputs
to integers and reduces the computation to GEMMs between integers. Below,
we refer to this FP4 version of the Ozaki scheme collectively as the
proposed method, and to its individual constructions as FP4 Ozaki schemes
I and II (abbreviated OzI-FP4 and OzII-FP4). The general formulation and
error analysis of Ozaki schemes I and II themselves are deferred to the
original papers~\cite{ozaki2012eftmm,ozaki2025schemeii}; this paper
confines itself to what the FP4 adaptation requires. The key is to encode
integers, without error, into sequences of values representable in E2M1.
Because the values representable in E2M1 are sparse and cannot represent
integers directly, this section introduces the base-13 FP4 limb
representation as that encoding. The representation provides the basic
operation of error-free GEMMs between small integers on FP4 Tensor Cores.

E2M1 is a 4-bit floating-point format with 1 sign, 2 exponent, and 1 mantissa bit; its value set is
$\lbrace 0, \pm 0.5, \pm 1, \pm 1.5, \pm 2, \pm 3, \pm 4, \pm 6 \rbrace$.
Doubling all values yields the integer set
$S = \lbrace 0, \pm 1, \pm 2, \pm 3, \pm 4, \pm 6, \pm 8, \pm 12 \rbrace$
(\figref{fig:coverage}).

In the proposed method, each coefficient $c_i \in S$ of an arbitrary
integer written positionally in base 13 is called a limb, after the
GMP~\cite{gmp} term for the digit-like
building block of multiple-precision integer arithmetic, and is stored as
one E2M1 value ($c_i/2$). Unlike the slices of prior work, which cut the
mantissa at bit boundaries, limbs are an arithmetic representation: a
sequence of remainders under division by 13. Prior work called units of
this kind slices in Ozaki scheme I and per-modulus residue components in
Ozaki scheme II; the term limb is not used there. The following lemma
gives the existence of this representation and the range representable
without gaps.

\begin{lemma}[Representation of all integers under shifts by multiples
of 13; gap-free range]\label{lem:crs}
Every integer $N$ can be written $N = c + 13n$ ($c \in S$, $n$ an
integer); that is, $S$ represents all integers under shifts by multiples
of 13 (only the residue classes $1$ and $12$ modulo 13 have two
representatives in $S$ and are representable redundantly). Consequently,
any integer can be decomposed by a greedy method (choosing at each limb
the residue representative of smallest absolute value) as
$N = \sum_{i=0}^{p-1} 13^i c_i$ ($c_i \in S$). Because $S$ is sparse,
however, the interval representable without gaps by $p$ limbs is
$|N| \le X_p = (13^p - 1)/3$ ($X_1 = 4,\ X_2 = 56,\ X_3 = 732, \dots$).
The proof is by induction: for $|N| \le X_p$, one can choose
$c_0 \equiv N \pmod{13}$ with $|c_0| \le 8$, and the quotient satisfies
$|(N - c_0)/13| \le X_{p-1}$.
\end{lemma}

Limb decomposition is an elementwise, branch-free recursion (take a
representative in $S$ of the remainder modulo 13, then recurse on the
quotient). That the greedy conversion generates only representatives with
$|c_i| \le 8$ strengthens the exactness conditions later.

Independently of the Ozaki scheme, this representation also yields a
bit-exact substitute for the integer GEMM of INT8 Tensor Cores
(INT8$\times$INT8$\to$INT32): the INT8 range is represented without error
by $p{=}q{=}3$ limbs ($X_3{=}732 \ge 128$), and the integer product is
reconstructed error-free from $9$ FP4 GEMMs. Ozaki schemes I and II of
this paper are built on this basic operation.

\section{FP4 Adaptation of Ozaki Scheme I}
\label{sec:ozaki1}

\begin{figure*}[tb]
\centering
\includegraphics[width=0.65\textwidth]{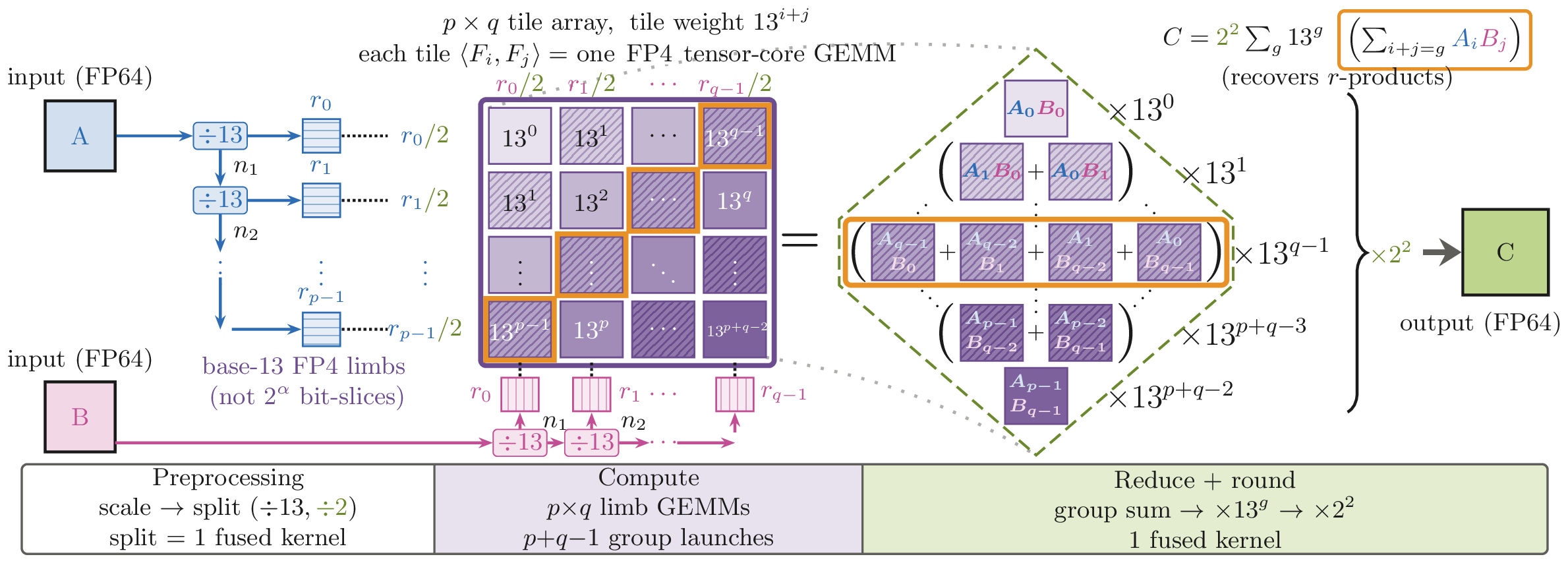}
\caption{Overall structure of OzI-FP4. The inputs are decomposed into
base-13 FP4 limbs by the $\div 13$ recursion; the $p \times q$
partial-product tiles (each corresponding to one FP4 Tensor Core GEMM)
are aggregated
along anti-diagonals ($i+j{=}g$) into $\sum_{i+j=g} A_i B_j$, and $C$ is
obtained with $\times 13^g$ and $\times 2^2$.}
\label{fig:hier1}
\end{figure*}

\figref{fig:hier1} shows the overall structure of OzI-FP4. The framework
inherits the error-free transformation of the Ozaki scheme, into which
the proposed method substitutes the base-13 FP4 limb representation. Below,
$\vec{A}, \vec{B}$ denote the inputs after integer conversion, whose
elements are integers; the conversion procedure and its accuracy are
described later in this section. Decomposing the operand vectors into
limbs as $\vec{A} = \sum_{i<p} 13^i \vec{a}_i$ and
$\vec{B} = \sum_{j<q} 13^j \vec{b}_j$, the inner product expands by
linearity into
\begin{equation}
\langle \vec{A},\vec{B} \rangle
= \sum_{i<p,\ j<q} 13^{i+j}\,\langle \vec{a}_i,\vec{b}_j \rangle .
\label{eq:expand}
\end{equation}
This generalizes the limb expansion shown for the INT8 substitute
($p{=}q{=}3$) in the previous section to arbitrary limb counts $p, q$.
Each $\langle \vec{a}_i, \vec{b}_j \rangle$ is an integer inner product
obtained by one FP4 MMA (one GEMM in the matrix-product case). Since the
MMA multiplies the stored values $\vec{a}_i/2, \vec{b}_j/2$, its output is
$\langle \vec{a}_i, \vec{b}_j \rangle / 4$, and a single $\times 2^2$ at
the end restores the integer product (\figref{fig:hier1}). The
coefficients $13^{i+j}$ are scaling factors outside the MMA, and terms
with the same $i+j$ can be grouped ($p+q-1$ groups). The final reduction
is a weighted sum over the $p{+}q{-}1$ groups, executed in a single
reduction kernel.

\begin{lemma}[Exactness condition]\label{lem:exact}
Limb values satisfy $|c| \le 12$, so products are at most $144$, and the
condition for a length-$K$ inner-product sum to stay within the exactly
representable integer range of FP32, without rounding, is
\begin{equation}
144K \le 2^{24} \quad (K \le 116{,}508).
\label{eq:exact1}
\end{equation}
The greedy conversion takes each limb as the representative of smallest
absolute value with $c \equiv x \pmod{13}$ ($|c| \le 6$) and replaces only
$\pm 5$, which is not in $S$, by $\mp 8$ in the same residue class, so the
generated limbs always satisfy $|c| \le 8$ (even for redundant
representations, e.g., $1 \equiv -12 \pmod{13}$, the side with the smaller
absolute value is chosen). The actual products are therefore $\le 64$,
and, accounting also for the group size $n_g$ (the number of $(i,j)$
pairs with $i{+}j{=}g$), $K \le 2^{18}/n_g$ holds ($K \le 17{,}476$ for
the deepest group $n_g{=}15$ at $p{=}q{=}15$).
\end{lemma}

\subsection{Limb Counts and the Number of GEMMs}
The required number of limbs is the smallest $p$ whose gap-free range
$X_p$ accommodates the mantissa width of the input. Because the
granularity is fine, $\log_2 13 \approx 3.70$ bits per limb, little is
wasted by allocating more limbs than necessary: INT8 fits in 3 limbs, the
FP32 mantissa (24 bits) in 7, and the FP64 mantissa (53 bits) in 15. The
smallest balanced configuration that exceeds the $2 \times 53 = 106$ bits
required for a product of two FP64 values is $15 \times 15$, which takes
$225$ GEMMs.

\subsection{Accuracy and Error Sources}
The accuracy of the proposed method has two layers: the computation is
exact with respect to the integer-converted inputs, and the accuracy with
respect to the original FP64 inputs is determined solely by the integer
conversion (quantization).
The 225-GEMM configuration with $p{=}q{=}15$ captures the 53-bit FP64
mantissa without truncation. This is, however, the capacity for the
mantissa of one element; elements whose exponents differ within a row
(column) undergo quantization by the shared exponent, so the effective
accuracy with respect to the original inputs can be lower.
The reduction reconstructing the result implements 128-bit integer
arithmetic in software as multiword arithmetic on four 32-bit words. GPUs
have no 128-bit integer instructions, but the reduction is $O(N^2)$,
lower order than the $O(N^3)$ of GEMM, and its share of the total runtime
including preprocessing is small: $14.1/7.3/3.9\,\%$ for
$N{=}4096/8192/16384$, decreasing as the problem grows. The output is
exact with respect to the integer-converted inputs
(Lemma~\ref{lem:exact}).
The accuracy with respect to the original FP64 inputs is determined solely
by the quantization introduced by the shared-exponent integer conversion.
This conversion uses a single scale exponent shared across each row of
$A$ and each column of $B$: $\tilde{A}_{i,:} = \mathrm{round}(A_{i,:} \cdot 2^{s_i})$,
where $s_i$ is chosen so that the converted value of the row maximum
$\max_j |A_{ij}|$ fits in the smaller of the range $X_p$
representable without gaps by $p$ limbs (Lemma~\ref{lem:crs}) and the
range $2^{53}$ in which integers are exact in FP64 (likewise for $B$ per
column).
Concretely, with the capacity $c = \lfloor \log_2 \min(X_p,\,2^{53})
\rfloor$ and $e_i$ satisfying $2^{e_i} \le \max_j |A_{ij}| < 2^{e_i+1}$,
we set $s_i = c - 1 - e_i$ (likewise $t_j$). If all elements of a row are
$0$, we set $s_i = 0$. The final output computes the integer matrix
product $\tilde{C} = \tilde{A}\tilde{B}$ and reconstructs
$C_{ij} = \tilde{C}_{ij}\,2^{-(s_i + t_j)}$ ($t_j$ is the column scale
exponent of $B$), rounded to FP64.
This conversion is common to the Ozaki schemes including those based on
FP8 and INT8, and the computation after conversion introduces no
additional error specific to FP4.

\section{FP4 Adaptation of Ozaki Scheme II}
\label{sec:ozaki2}

\begin{figure}[tb]
\centering
\includegraphics[width=\linewidth]{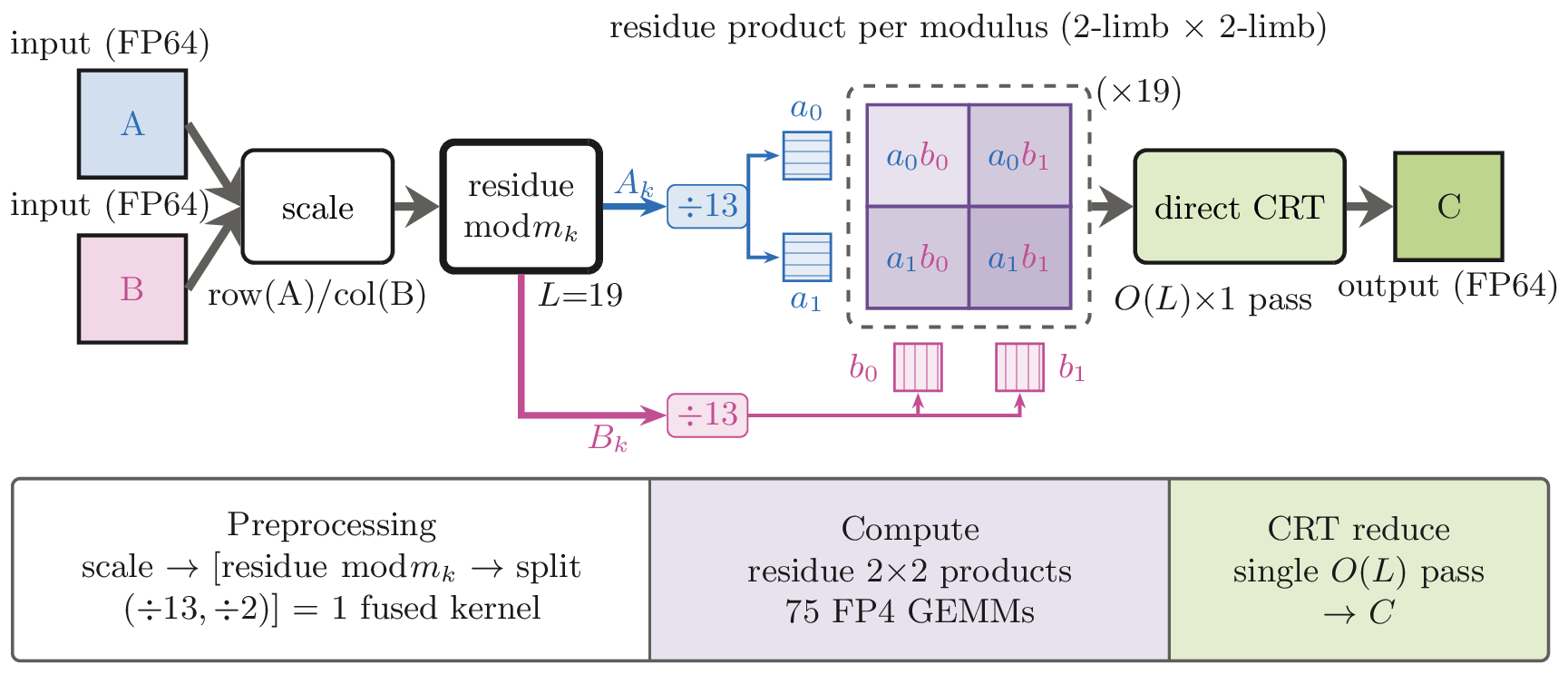}
\caption{Overall structure of OzII-FP4 (same format as
\figref{fig:hier1}). After scaling, limb decomposition including the
residue computation mod $m_k$ runs in one fused kernel; after a total of
75 FP4 GEMMs over the $L{=}19$ moduli, a single-pass direct CRT
reconstructs the result.}
\label{fig:hier2}
\end{figure}

\figref{fig:hier2} shows the overall structure of OzII-FP4.
Ozaki scheme II~\cite{ozaki2025schemeii} computes the product in parallel
as residues modulo pairwise coprime moduli $m_1, \dots, m_L$ and
reconstructs it without error by the Chinese remainder theorem (CRT)
(GEMMul8~\cite{uchino2026fp8ozakiii,riken2026gemmul8} is the FP8/INT8
implementation). Whereas the number of GEMMs of Ozaki scheme I grows with
the product $p \times q$ of the limb counts, Ozaki scheme II incurs a fixed
cost per modulus, growing only linearly in the number of moduli, and
therefore becomes advantageous in the high-accuracy regime.

\subsection{Representability Condition and Modulus Selection}
\label{sec:moduli}
The key to the FP4 adaptation is a representability condition: every
residue class of a modulus $m$ must contain a representative expressible
with two limbs. Two limbs represent the range $X_2 = 56$ without gaps
(Lemma~\ref{lem:crs}), so every modulus $m \le 113$, whose representative
of smallest absolute value falls within $\pm 56$, necessarily satisfies
it. Exhaustive
verification further finds four exceptions above this bound,
$\lbrace 115, 117, 143, 169 \rbrace$, for which every residue class is
still representable with two limbs (by representatives not necessarily
of smallest absolute value); the two-limb moduli are thus ``all integers
$m \le 113$'' plus these four exceptions. For example, for $m{=}115$ the
residue class $57$ is not representable in two limbs by its
smallest-magnitude representative $57$, but is by the representative
$-58 = 13 \cdot (-4) - 6$ of the same class. Adopting $169 = 13^2$
excludes the other multiples of 13 ($117, 143$, etc.), which are not
coprime with 169. Taking only 169 among the multiples of 13 and choosing
greedily in descending order subject to pairwise coprimality yields the
$L{=}19$ moduli
\begin{align*}
\{&169,\, 115,\, 113,\, 112,\, 111,\, 109,\, 107,\, 103,\, 101,\, 97,\\
&89,\, 83,\, 79,\, 73,\, 71,\, 67,\, 61,\, 59,\, 53\}
\end{align*}
(composite numbers included). We write the product of the moduli as
$P = \prod_i m_i \approx 2^{123}$ from here on. Each element of the matrix
product is a signed sum of $K$ integer products (after conversion each
element is an integer of absolute value $< 2^{53}$, so each product is
$< 2^{106}$), so its absolute value is below $K \cdot 2^{106}$. The signed
unique-reconstruction condition is therefore
\begin{equation}
2K \cdot 2^{106} < P \approx 2^{123},
\label{eq:crt}
\end{equation}
which is amply satisfied for $K \le 16384$, where the left side is at most
$2^{121}$. Exploring all combinations of 18 pairwise coprime moduli from
the representable candidates yields a product of at most $2^{117.80}$; no
set of 18 satisfies $P > 2^{121}$, so $L{=}19$ is minimal.
For each modulus $m$, the residues of both operands are represented with
two limbs, $a \equiv a_0 + 13a_1,\ b \equiv b_0 + 13b_1 \pmod{m}$, and
their product is composed from the following four partial products:
\[
ab \equiv a_0 b_0 + 13(a_0 b_1 + a_1 b_0) + 13^2 a_1 b_1 \pmod{m}.
\]
Each modulus takes the four pairwise products between the two residue
limbs of the operands; only for $169 = 13^2$ does the last term
$a_1 b_1$ vanish because $13^2 \equiv 0 \pmod{169}$, saving one GEMM for
a total of $4L - 1 = 75$. In
addition, for the seven moduli $\lbrace 83, 73, 71, 67, 61, 59, 53
\rbrace$ that can represent the sum limbs $(c_0{+}c_1)$, Karatsuba's
method can further reduce each to three (68 in total). In contrast to the
FP8 version of the existing implementation GEMMul8, which reduces every
modulus to three products using Karatsuba's method (7 moduli) and
perfect-square identities ($1089 = 33^2$, etc.; 6 moduli), the FP4
representation lacks the range margin these require, so the reduction
applies only partially. Under kernel fusion, moreover, this reduction in
GEMM count does not translate into speed, and we do not use it in our
evaluation; the reason is given in Section~\ref{sec:ozii-opt}.

\begin{lemma}[Exactness conditions, Ozaki scheme II]\label{lem:exact2}
Residue limbs also satisfy $c_i \in S$ ($|c_i| \le 12$), so products are
at most $144$. The upper bounds on the inner-product length $K$ for exact
computation follow from four conditions, requiring the values of each
stage to stay within its representable range:
{\small\setlength{\arraycolsep}{3pt}
\[
\begin{array}{lrcll}
\text{each partial product} & 144K & \le & 2^{24} & (K \le 116{,}508)\\
\text{cross-term accumulation} & 288K & \le & 2^{24} & (K \le 58{,}254)\\
\text{residue composition (INT32)} & 28{,}224K & \le & 2^{31} & (K \le 76{,}087)\\
\text{CRT unique reconstruction} & 2K \cdot 2^{106} & < & P & (K \le 76{,}549)
\end{array}
\]
}
The first is identical to Lemma~\ref{lem:exact}; the second is the bound
$288K$ of the implementation that accumulates the two cross terms sharing
weight $13$ in a single accumulator (Section~\ref{sec:ozii-opt}); the
third is the condition that the residue composition performed at the GEMM
output stage, $(1 + 2{\cdot}13 + 169) \cdot 144K = 28{,}224K$, fits in
INT32; the fourth is the CRT unique-reconstruction condition
\eqref{eq:crt}. The overall limit is therefore set by the cross-term
accumulation condition, $K \le 58{,}254$, which our evaluation at
$K \le 16{,}384$ satisfies with a margin of about $3.6\times$.
\end{lemma}

\subsection{Exact Single-Pass Direct CRT Reconstruction in Integer
Arithmetic}
Reconstruction from a residue system involves a trade-off between
exactness and cost. Garner's algorithm, the classical standard for exact
reconstruction, determines the digits of a mixed-radix representation
successively from the residues of each modulus and thus requires $O(L^2)$
residue updates for $L$ moduli.
The direct CRT, which evaluates the contributions of all moduli at once
as a weighted sum, costs only $O(L)$, and the original paper on Ozaki
scheme II also formulates the reconstruction in this form. The
reconstruction of the existing implementation GEMMul8 is likewise a
direct CRT, but it keeps the precomputed weights in FP64 (in two-word
double-double format in configurations where $P$ exceeds the 53-bit
mantissa) and accumulates them in floating point. Our implementation, in
contrast, performs this direct CRT in integer arithmetic, keeping the
reconstruction exact at $O(L)$ cost.

Concretely, from the precomputed 128-bit integer weights $w_i$ (with
$M_i{=}P/m_i$, $w_i = M_i\,(M_i^{-1} \bmod m_i)$) and the residue products
$r_i$ ($AB \bmod m_i$) of each modulus, we evaluate
$x = \sum_i r_i w_i \bmod P$ in a single pass. The sum
$S = \sum_i r_i w_i$ is accumulated without error in 160-bit integers,
and the quotient $\lfloor S/P \rfloor$ is estimated in FP64 and
determined exactly with a $\pm 1$ correction.\footnote{The quotient $\lfloor S/P \rfloor$
never exceeds $\sum_i (m_i - 1) = 1753$ because $S < (\sum_i r_i) P$, and
its maximum in this configuration is $754$. The relative error of
converting $S$ to FP64 and dividing by $P$ stays within a few multiples
of the unit roundoff $u = 2^{-53}$, so the absolute error of the quotient
is on the order of $10^{-12}$ and the $\pm 1$ correction determines the
correct quotient. Moreover $S < 2^{133}$, so the $160$-bit accumulation
cannot overflow.} Mapping the resulting $x$ to its representative in the
symmetric range ($x - P$ if $x > P/2$) reconstructs the integer-converted
product bit-exactly under the unique-reconstruction condition
\eqref{eq:crt}. Finally, the result is converted to FP64 with a single
round-to-nearest operation. In our implementation, these 128- and 160-bit integer
operations are carried out as multiword software based on 32-bit integer
arithmetic, but this reduction is $O(N^2)$, lower order than the
$O(N^3)$ of GEMM, and occupies a small share of the runtime.

\section{Implementation and Optimization}
\label{sec:impl}

\subsection{Implementation and Evaluation Environment}
\label{sec:env}
Implementation and evaluation are carried out on an NVIDIA RTX PRO 6000
Blackwell Workstation Edition (sm\_120). The nominal Tensor Core peaks
(dense) are 2000 TFLOPS for FP4, 1000 TFLOPS for FP8, and 1000 TOPS for
INT8; FP4 has twice the throughput of FP8/INT8. FP64 is not listed in the
datasheet and measures $1.85$ TFLOPS with cuBLAS. Memory bandwidth is
1792 GB/s nominal and 1490 GB/s as measured by a copy benchmark.
We use GPU driver 595.71.05, CUDA~12.8, PyTorch~2.11.0, and Triton~3.6.0.
For comparison, we use the distributed version 3.1.0 of GEMMul8,
unmodified, built with the CUDA~13.2 nvcc and \texttt{-std=c++20 -O3
-gencode arch=compute\_120,code=sm\_120}. Time is measured with CUDA
events: after 3 warm-up runs, we take 12 measurements ($38$ for
$N{=}8192$) and report the median. Host--device transfers are not included.

OzI-FP4 and OzII-FP4 are implemented entirely as GPU kernels in
Triton~\cite{tillet2019triton}
on PyTorch. Triton is a DSL in which GPU kernels are written tile by
tile in Python while the compiler automates low-level optimization such
as memory coalescing, shared-memory placement, and register allocation.

\subsection{Common Preprocessing Optimizations}
OzI-FP4 and OzII-FP4 share their preprocessing and kernel-fusion
strategy. Fusion appears in three places: the fused limb decomposition,
which combines the preprocessing steps of limb decomposition and E2M1
encoding into one kernel; the epilogue fusion (described below),
which combines the residue GEMMs of all moduli of OzII-FP4 (the matrix
products of the residues for each modulus) and the residue composition
into one kernel; and the single-kernel reduction.

Preprocessing consists of two steps: scaling by shared exponents and the
fused limb decomposition. The latter performs the $\div 13$ recursion and
the E2M1 encoding in one fused kernel, reading the integer-converted
input only once and writing out the limb planes (the per-limb coefficient
matrices) in a single sweep (for OzII-FP4 the kernel also includes the
residue computation). Output tiles are kept small, at $16$ rows, so the working registers of the decomposition recursion do not spill (tiled
encoding). Each limb GEMM achieves the highest performance on the
Blackwell-generation FP4 Tensor Cores when its operands are arranged in
what BLAS convention calls the TN form ($A^{\mathsf T}B$, with $A$
transposed and $B$ untransposed, aligning the contiguous dimension of
both operands with the contraction dimension $K$). The $B$-side limb planes
are therefore written out directly in the $K$-contiguous orientation via
in-register transposes, so no separate transpose pass exists.

\subsection{OzI-FP4: Effect of the Optimizations}
OzI-FP4 applies the common optimizations of the previous subsection
(fused limb decomposition, tiled encoding, direct TN write-out)
unchanged. The
fused limb decomposition is a kernel that writes out all $p$ limb planes
in one sweep, and outperforms a non-fused implementation with one
kernel per limb plane by $10.0\times$ in kernel time alone. The
end-to-end runtime (including preprocessing) for 53-bit mantissas (FP64,
$p{=}q{=}15$) is reduced by a factor of $2.9$ over the three optimization
stages (\figref{fig:breakdown}), and with the preprocessing shortened,
the overall pipeline becomes GEMM-bound.

\begin{figure}[tb]
\centering
\includegraphics[width=\linewidth]{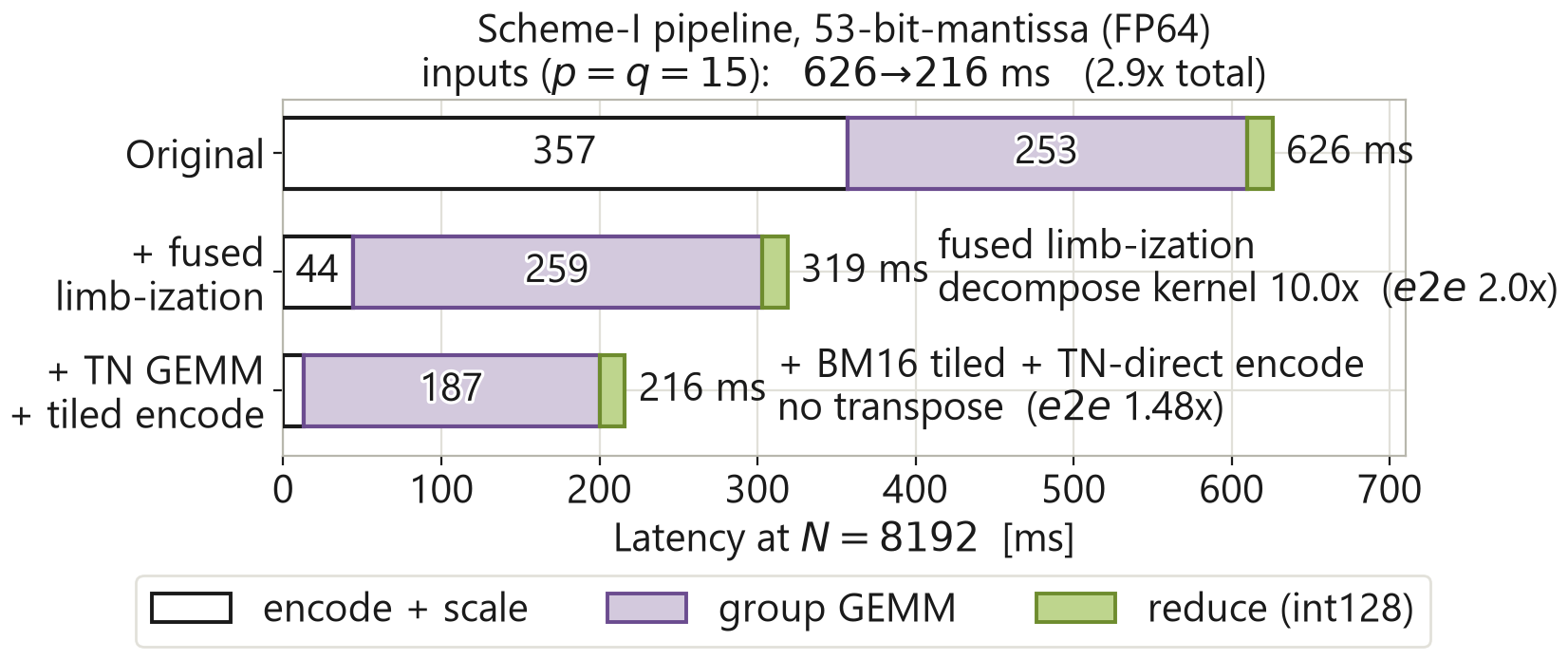}
\caption{Three optimization stages of the OzI-FP4 pipeline (FP64 input
with a 53-bit mantissa, $p{=}q{=}15$, $8192^3$). The bar of each stage
shows the breakdown into preprocessing (scaling, limb decomposition,
encoding), the GEMM part, and the reduction.}
\label{fig:breakdown}
\end{figure}

\subsection{OzII-FP4: Effect of the Optimizations}
\label{sec:ozii-opt}
For all 19 moduli, the partial products are grouped by weight $13^{i+j}$
and computed in one kernel (one launch), with the residue composition
embedded in the epilogue, the write-out stage of the GEMM. This form of
kernel fusion is called epilogue fusion in CUTLASS and similar
libraries; we use this term below. Each thread block loads four operand planes
($128 \times 128$ tiles of $a_0, a_1, b_0, b_1$, stepping $128$ along
$K$; the largest configuration fitting within the shared-memory limit)
exactly once, keeps three accumulators per modulus in registers
($a_0 b_0$; the sum of the cross terms $a_0 b_1 {+} a_1 b_0$, sharing
weight $13$; and $a_1 b_1$), and completes the residue composition in the
epilogue without materializing per-group intermediate matrices. The
final term $a_1 b_1$, which vanishes for modulus $169 = 13^2$
(Section~\ref{sec:moduli}), is also computed the same way as the other
moduli to avoid branching and canceled in the composition, so
the implementation unconditionally executes
$4L = 76$ GEMMs (the model of Section~\ref{sec:model} uses the mathematically
required $75$). Because kernel-launch overhead and memory accesses are
reduced, the compute stage (GEMM and reconstruction, excluding
preprocessing) outperforms a per-modulus-launch configuration using Karatsuba's method
($68$ GEMMs). Under fusion, however, the additional
loads of the two sum-limb planes increase shared-memory pressure and the
reduction in GEMM count does not translate into speed, so our
implementation did
not use Karatsuba's method. This partial Karatsuba may become effective
on future architectures with more shared memory. Replacing the $O(L^2)$
reduction based on Garner's algorithm, the standard for exact
reconstruction, with the direct $O(L)$ CRT while keeping exactness, and
then combining the residue GEMMs and residue composition of all moduli
into one launch by epilogue fusion, reduces the end-to-end runtime by a
factor of $3.1$ (\figref{fig:breakdownii}).

\begin{figure}[tb]
\centering
\includegraphics[width=\linewidth]{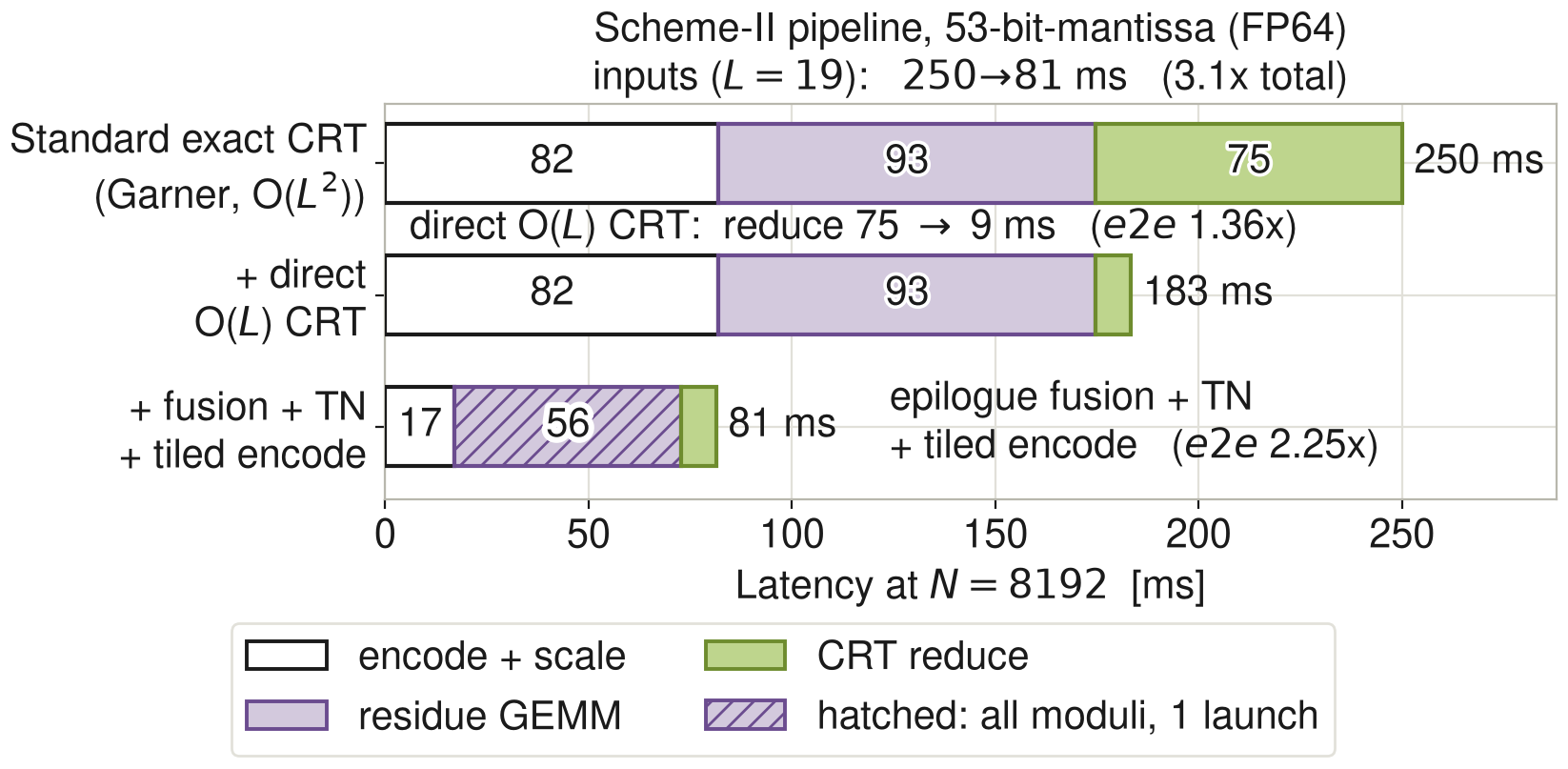}
\caption{Three optimization stages of the OzII-FP4 pipeline (FP64 input
with a 53-bit mantissa, $L{=}19$, $8192^3$). The stages show the
breakdown by reduction method (from Garner's algorithm, the standard for
exact reconstruction, to the direct CRT) and by kernel fusion.}
\label{fig:breakdownii}
\end{figure}

\section{Evaluation}
\label{sec:exp}

\begin{table}[tb]
\centering
\caption{Performance comparison in FLOPS (equivalent-DGEMM TFLOPS
$= 2N^3/t$; higher is faster). ``Comp.'' = compute stage; ``Total'' =
end-to-end including preprocessing.}
\label{tab:flops}
\small
\setlength{\tabcolsep}{2.2pt}
\begin{tabular}{@{}lrrrrrr@{}}
\hline
& \multicolumn{2}{c}{$4096^3$} & \multicolumn{2}{c}{$8192^3$} & \multicolumn{2}{c}{$16384^3$} \\
\cmidrule(lr){2-3}\cmidrule(lr){4-5}\cmidrule(lr){6-7}
Method & Comp. & Total & Comp. & Total & Comp. & Total \\
\midrule
cuBLAS DGEMM & 1.85 & 1.85 & 1.85 & 1.85 & 1.85 & 1.85 \\
OzI-FP4 & 5.60 & 4.96 & 5.41 & 5.09 & 5.58 & 5.41 \\
OzII-FP4 & 16.28 & 10.81 & 17.11 & 13.52 & 17.31 & 15.26 \\
GEMMul8-FP8 & 13.68 & 11.64 & 15.41 & 13.88 & 15.68 & 14.65 \\
GEMMul8-INT8 & 32.41 & 25.45 & 39.78 & 33.49 & 40.94 & 36.39 \\
\bottomrule
\end{tabular}
\end{table}

The main comparison of this section involves five methods: the proposed
OzI-FP4 and OzII-FP4; the FP8 version of Ozaki scheme II in the existing
implementation GEMMul8 (GEMMul8-FP8) and its INT8 version
(GEMMul8-INT8); and native cuBLAS DGEMM as the baseline. These
abbreviations are used below.
GEMMul8 has two scaling modes, one favoring accuracy and one favoring
speed; this paper compares against the default accuracy-oriented mode.

\subsection{Accuracy Evaluation}
\label{sec:acc}

Accuracy is compared at each method's optimal configuration for DGEMM
emulation, that is, the smallest limb and modulus counts that capture the
FP64 mantissa without error. These minimal configurations are the $13$
moduli of~\cite{uchino2026fp8ozakiii} for FP8, and $L{=}19$ moduli for
OzII-FP4 and $p{=}q{=}15$ limbs for OzI-FP4 for the proposed method. That
the proposed method needs more moduli and limbs than the $13$ of FP8 is
due to fewer bits per FP4 limb.

Each element of the test matrices for the accuracy evaluation is
generated as $\pm(1+u) \cdot 2^{e}$ ($u$ a uniform random number in $[0,1)$, $e$ a
uniform integer in $\lbrace 0, \dots, \varphi \rbrace$). The element magnitudes are distributed roughly from $1$ to $2^{\varphi}$.
Sweeping $\varphi$ evaluates the accuracy for inputs with different
spreads of magnitude.
The error is obtained by taking, for each output element, the difference
from the result of $220$-bit arithmetic, normalizing it by the sum of absolute values of the multiplied terms
$\sum_k \vert A_{ik}\vert\,\vert B_{kj}\vert$, and taking the maximum
over all elements~\cite{fasi2023multiword}. The inner-product length is
$8192$, the middle size of the speed evaluation, and all $16{,}384$
elements of a $128 \times 128$ output are evaluated. All methods are
measured on the same matrices with the same metric.

\figref{fig:g16acc} shows the accuracy results. All emulations are more
accurate than cuBLAS DGEMM, and the errors of the proposed method and
GEMMul8-FP8 are of the same order. In effective bits (the error
$\epsilon$ expressed as $-\log_2\epsilon$), the proposed method is up to
$0.9$ bit above for
$\varphi \le 4$ and up to $1.5$ bits below for $\varphi \ge 8$. The
proposed method declines faster as $\varphi$ grows, dropping $5.9$ bits
from $58.2$ bits at $\varphi{=}0$ to $52.3$ bits at $\varphi{=}32$
($3.5$ bits for the FP8 version). This corresponds to the shared-exponent
integer conversion losing more bits for elements whose exponents lie
farther apart within a row. At $\varphi{=}0$, however, it reaches
$3.1 \times 10^{-18}$, the minimum error corresponding to a single final
rounding, and there it surpasses GEMMul8-FP8. This corresponds to the
fact that the only error source of the proposed method is the integer
conversion (shared-exponent quantization): at $\varphi{=}0$ the
quantization too is error-free and the product of the original inputs is
obtained with a single final rounding, whereas the larger $\varphi$ is,
the larger the quantization error of the small elements.
The proposed method remains exact with respect to the converted integer
inputs. The curves of OzI-FP4 and OzII-FP4 coincide exactly because both
compute the integer product of the same converted inputs without error:
their constructions differ (positional limb expansion versus RNS and
CRT), but both compute this integer product exactly and round only once
at the end, so the outputs agree bit for bit.

\begin{figure}[tb]
\centering
\includegraphics[width=0.75\linewidth]{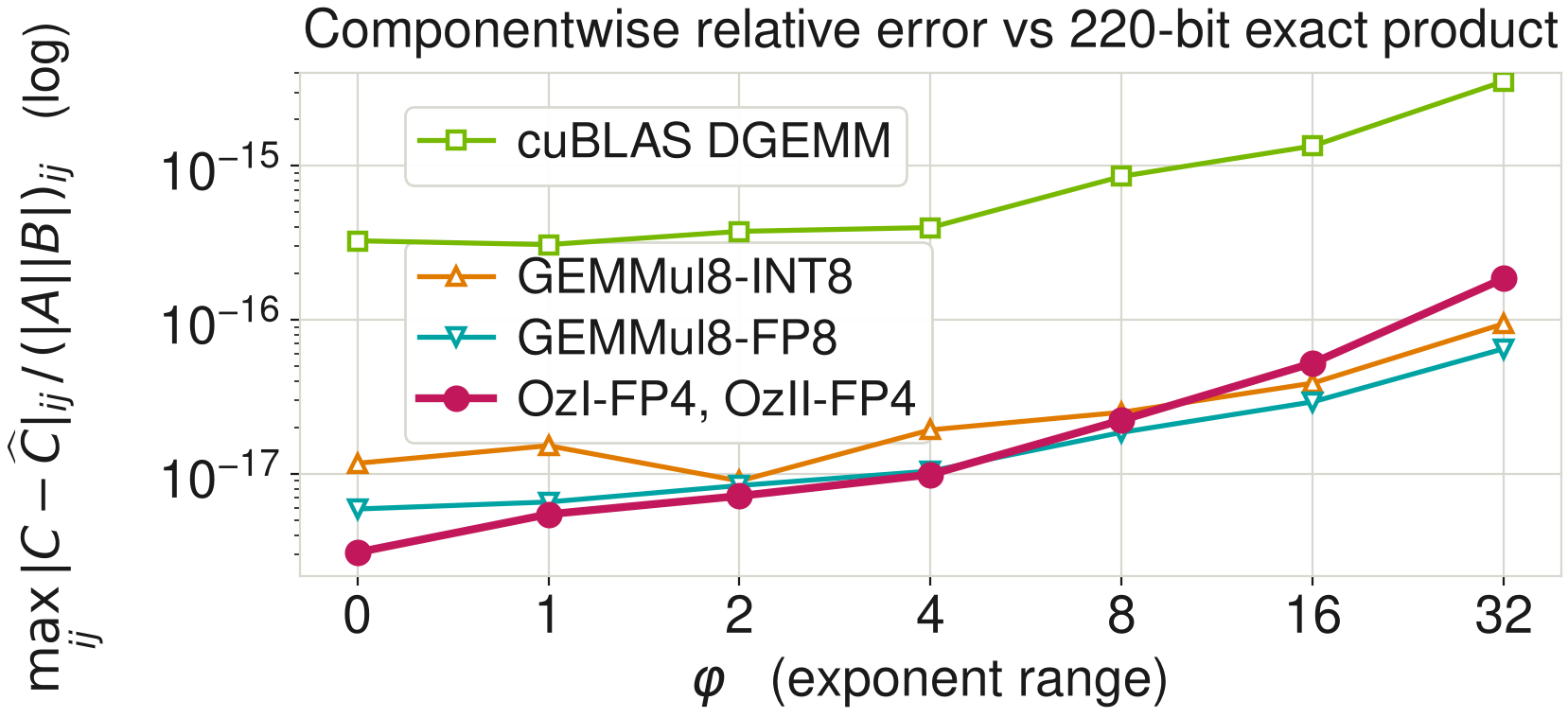}
\caption{Accuracy comparison (inner-product length $8192$, all elements
of a $128 \times 128$ output). The curve of the proposed method is
shared by the two configurations (OzI-FP4 and OzII-FP4), whose outputs
agree bit for bit.}
\label{fig:g16acc}
\end{figure}

\subsection{Performance Evaluation}
\label{sec:duel}

The environment and timing methodology are those of
Section~\ref{sec:env}. \tabref{tab:flops} shows the performance. We
report performance for two scopes: the compute stage, which excludes the
preprocessing into low precision, and the end-to-end execution, which
includes it.
On this GPU, which has INT8 units of the same throughput as FP8,
GEMMul8-INT8 is the fastest. In the generation where the INT8 units are
reduced (B300), however, and in the Rubin generation expected to share
that trend, GEMMul8-FP8 versus OzII-FP4 becomes the relevant comparison.
OzII-FP4 outperformed GEMMul8-FP8 in the compute stage by
$1.10$--$1.19\times$ across the whole range ($N{=}4096/8192/16384$). This
advantage reflects two contributions: the implementation optimization in
which the epilogue fusion of the residue GEMMs (computing all 19 moduli
and all groups in one kernel and one launch without materializing
per-group intermediate matrices) removes the bottleneck of kernel-launch
overhead and attains a high fraction of peak performance, and the
theoretical advantage determined by the FP4:FP8 throughput ratio and the
GEMM-count ratio. The following subsection quantifies the two
contributions. End
to end as well, with the preprocessing shortened by about $5\times$
through tiled encoding, the end-to-end time of OzII-FP4 is $0.96\times$
that of GEMMul8-FP8 at $16384^3$, shorter and thus faster, and at
$4096^3/8192^3$ it takes $1.08/1.03\times$ the time, only slightly
slower. Against cuBLAS DGEMM it is $8.8/9.2/9.4\times$ faster in the
compute stage and $5.8/7.3/8.2\times$ end to end.

\subsection{Performance Model and Validation Against Measurements}
\label{sec:model}

The advantage of OzII-FP4 over GEMMul8-FP8 shown in the previous
subsection can be explained quantitatively by a runtime model. The
proposed method's runtime can be modeled with two terms, the compute-bound GEMM
part $t_{\mathrm{GEMM}}$ and the memory-bound preprocessing
$t_{\mathrm{pre}}$:
\begin{equation}
t_{\mathrm{total}} \approx t_{\mathrm{GEMM}} + t_{\mathrm{pre}}
= \frac{2QN^3}{R} + \frac{D}{\beta},
\label{eq:model}
\end{equation}
where $Q$ is the number of GEMMs, $R$ the effective peak of the FP4
Tensor Cores, $D$ the bytes moved, and $\beta$ the effective memory
bandwidth. The GEMM part is $O(N^3)$, while the preprocessing and the
reconstruction reduction are both lower order at $O(N^2)$, so the GEMM
part dominates at large $N$. Below, $t_{\mathrm{GEMM}}$ plus the
theoretical reduction time is called the theoretical compute-stage time.
The effective values on this GPU are $R = 1408$ TFLOPS (single-kernel
measurement of our Triton kernel) and $\beta = 1490$ GB/s (copy
benchmark).

\begin{figure}[tb]
\centering
\includegraphics[width=0.7\linewidth]{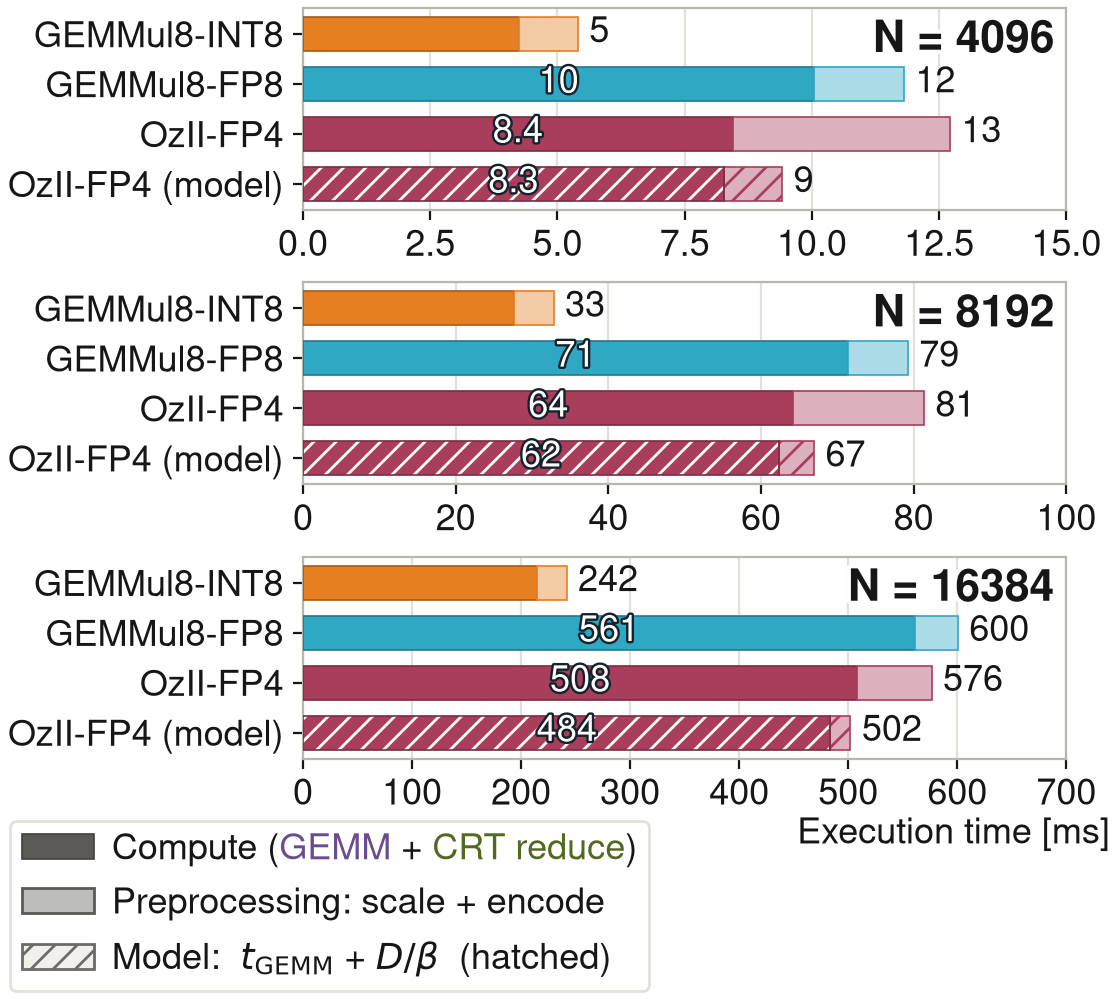}
\caption{Breakdown of the end-to-end time of the three FP64 emulations
(GEMMul8-INT8, GEMMul8-FP8, and OzII-FP4) for each $N$. Dark bars show
the compute stage; light bars show preprocessing. For OzII-FP4, the
model (hatched) is placed directly beneath the measurement.}
\label{fig:g16speed}
\end{figure}

The measurements follow this model \eqref{eq:model}.
\figref{fig:g16speed} shows the runtime breakdown of each method into
compute stage and preprocessing, together with the theoretical values
($t_{\mathrm{GEMM}}$, $D/\beta$) for OzII-FP4. The compute stage
(GEMM and the reconstruction reduction) exceeds the
theoretical value including the reduction by $2/3/5\,\%$ for
$N{=}4096/8192/16384$, respectively (the reduction is lower order at
$O(N^2)$, so the larger $N$ is, the closer the measurement approaches
$t_{\mathrm{GEMM}}$ itself). The preprocessing is about $3.7\times$ the
bandwidth bound determined by its bytes moved (the modular arithmetic of
the encoding adds overhead on top). The runtime at large $N$ is therefore
predicted by $t_{\mathrm{GEMM}}$, proportional to the GEMM count $Q$.
Note that this agreement is relative to $R$, the effective peak of
our own kernel, and is a separate matter from the absolute utilization of
the arithmetic units.

The model explains the performance ordering through
arithmetic-throughput ratios: for the same accuracy target, method $X$
($Q_X$ GEMMs at rate $R_X$) outperforms method $Y$ when
$Q_X/R_X < Q_Y/R_Y$. For example, OzI-FP4 takes asymptotically about
$3.6\times$ the GEMMs of the INT8 slice approach, so it becomes
advantageous on hardware whose FP4:INT8 throughput ratio exceeds that
factor. Within this framework, the advantage of OzII-FP4
over the existing implementation GEMMul8-FP8 separates into two stages.

The first stage is the ordering determined only by the peak throughput
of each arithmetic unit and the number of GEMMs, independent of
implementation optimization. Assume both methods realize their
respective nominal peaks. OzII-FP4 has $Q{=}75$; GEMMul8-FP8 has
$Q{=}39$ for the 53-bit FP64 mantissa ($13$ moduli at $3$ products per
modulus); and under the FP4:FP8 throughput ratio of $2$ (nominal
$2000/1000$), the advantage, as a theoretical value based on the
mathematically required GEMM counts, is a factor of
$39 \times 2 / 75 = 1.04$. In terms of the GEMMs the
implementations execute, GEMMul8-FP8 executes $39$ (the one scaling GEMM
is executed on the preprocessing side) and OzII-FP4 executes $76$
including the canceled term; on that basis the factor is
$39 \times 2/76 = 1.03$, but the definition of the attained fraction of
peak performance shifts correspondingly and the explanation of the
measured ratio is unchanged.
FP4 has twice the rate but needs about $1.9\times$ the GEMMs because of
its smaller mantissa, so the throughput gain is largely offset (break-even at a required throughput ratio of $1.9$). The value
still exceeds $1$, so FP4 beats FP8 in the compute stage even with
implementation differences removed. This is the theoretical advantage
net of implementation differences.

The second stage is the difference in how much of its own peak each
implementation attains as a result of implementation optimization (the
attained fraction of peak performance). OzII-FP4 nearly realizes its own
effective peak ($R$) through the epilogue fusion of the residue GEMMs
and operand reuse, whereas GEMMul8-FP8 does not reach its own peak.
Relative to the nominal peaks, the measured attained fractions of peak
performance (based on the mathematically required counts) are
$61$--$65\,\%$ for OzII-FP4 and $53$--$61\,\%$ for GEMMul8-FP8, whose
ratio is $1.06$--$1.14\times$.

The product of the $1.04$ factor determined by the throughput and
GEMM-count ratios and the $1.06$--$1.14$ ratio of attained fractions of
peak performance explains the measured $1.10$--$1.19\times$ advantage
(previous subsection) almost exactly. The
observed advantage is not attributable to either factor alone. Applying
Karatsuba's method to the seven moduli of Section~\ref{sec:moduli} would
reduce $Q$ to $68$, for a theoretical advantage of $1.15$, but under
epilogue fusion it is counterproductive and we do not adopt it. GEMMul8 is used as
distributed, targeting multiple GPUs, without tuning to this
GPU; the gap in the attained fraction of peak performance above also
includes this difference in maturity.

\section{Conclusion}

By exploiting the integer nature of FP4 (E2M1) to represent
arbitrary integers without error in base-13 FP4 limbs, and by keeping
intermediate sums in the integer range of FP32 accumulators, this
work constructed error-free integer GEMM on FP4 Tensor Cores. On this
foundation, we adapted both Ozaki schemes I and II, which prior work had
carried only as far as FP8, to FP4, and provided a bit-exact emulation of
INT8 Tensor Cores by the same mechanism.
In measurements on an RTX PRO 6000 Blackwell, OzII-FP4 kept accuracy at
or above the DGEMM level while performing on par with GEMMul8, the
existing FP8-based implementation of Ozaki scheme II, and surpassing it
at problem size $16384^3$. The performance advantage over GEMMul8-FP8 was
explained quantitatively by the runtime model as the product of the
theoretical difference stemming from the FP4:FP8 throughput ratio and the
GEMM-count ratio, and the higher attained fraction of peak performance
from the implementation optimizations, most notably the epilogue fusion
of the residue GEMMs. This work makes FP4 Tensor Cores usable for
scientific computing. The source code of the proposed method is publicly
available on GitHub.\footnote{\url{https://github.com/FP4-is-All-you-Need/Oz-FP4}}

\section*{Acknowledgment}
This work was supported by JSPS KAKENHI Grant Number JP25K24387. Part of
this work was supported by the Joint Usage/Research Center for
Interdisciplinary Large-scale Information Infrastructures (JHPCN) in
Japan (Project ID: jh260065).

Generative AI assistants (Anthropic Claude and OpenAI Codex) were used
under the authors' direction to draft and translate portions of the
manuscript text, to write the figure-plotting scripts, and to assist in
developing the experimental code. All methods, results, and claims were
designed, verified, and approved by the authors, who take full
responsibility for the content of this paper.

\bibliographystyle{IEEEtran}
\bibliography{references}

\end{document}